\documentclass[12pt]{article}
\usepackage[dvips]{graphicx}
\usepackage{amsfonts,amsmath,amsthm,amssymb}
\usepackage{mathrsfs}

\author{Leonardo Ort\'{i}z\footnote{leonardoortizh@gmail.com}, Marcelo Amaral and Klee Irwin\\
Quantum Gravity Research, Los Angeles, CA, U.S.A}

\title{Aspects of aperiodicity and randomness in theoretical physics}

\begin{document}

\maketitle

\begin{abstract}
In this work we explore how the heat kernel, which gives the solution to the diffusion equation and the Brownian motion, would change when we introduce quasiperiodicity in the scenario. We also study the random walk in the Fibonacci sequence. We discuss how these ideas would change the discrete approaches to quantum gravity and the construction of quantum geometry.
\end{abstract}
\vspace{1cm}


\newpage

\section{Introduction} 

The idea of randomness is very old and studies of it date back to Laplace in the eighteenth century. Nowadays with the invention of quantum mechanics probability plays a fundamental role in our civilization. In particular it is expected that the theory of quantum gravity will have also probability as an important element in its structure. On the other hand, geometry is well known to be fundamental in modern theories of physics, in particular, general relativity is a geometrical theory. So it is natural that its quantum version will also have geometry as a fundamental element. In this work we explore how these two elements play a role in some aspects of theoretical physics. Additionally we introduce the concept of aperiodicity in some parts of this work and see how this introduction  would modify, even if mild, the theoretic results. 

In this paper we discuss the notion of quantum geometry \cite{am97} using tools from quasicrystals and quasiperiodic functions \cite{baa13} as an alternative to canonical quantization studied in \cite{am19}, mainly the modifications of quantities of interest, such as the heat kernel and entropy, due to the introduction of quasiperiodicity in the setting. Although these modifications seem to be mild, they have the potential to be important in derivations of thermodynamical quantities, which we plan to develop in the future. We hope that this work gives some input on the interface between probability and geometry.

In \cite{am97} a kind of quantum geometry is constructed in one and two dimensions. This is done using random walks and random two-dimensional surfaces. This is done with the idea of obtaining a quantum description of gravity, at least in two dimensions. This is limited in several aspects with the main limitation that a realistic theory of gravity should be four-dimensional. The main idea of the present work is to make something similar to \cite{am97} but not only with random walks and random two-dimensional surfaces but also with quasiperiodic trajectories as described in \cite{baa13}. In this manuscript we will consider one and two dimensional quasiperiodic trajectories but one of our goals is to make it with quasiperiodic trajectories inherited from quasicrystals in several dimensions. Also once we have this under control we will try to not only have nonperiodicity but also stochasticity. In the near future we will try to do what is done with the Polyakov action in \cite{jon99} but this time with the Einstein-Hilbert action. Also instead of letting the lattice size go to zero we will probe the geometry with a massive particle so we can feel the granular structure of spacetime. 

In this context the quasiperiodicity will be given by the trajectory of the quasiparticle. It is important to note that in this step we can introduce nonperidicity and stochasticity.

In the scenario we are describing here we would like to obtain the analogous to the Einstein field equations as something emergent in the same spirit as thermodynamics is obtained from the microscopic statistical mechanics.  

This manuscript is organized as follows: 
In section \ref{seckernel} we describe our approach to the build up of quantum geometry with a discussion of the heat kernel and quasiperiodicity. 
In section \ref{secwalk} we study the random walk in a Fibonacci chain in one dimension and its generalization to two dimensions. 
In section \ref{secentropy} we discuss further ideas on partition function and entropy. 
In section \ref{secbtz} we discuss a concrete example of the entropy of the BTZ black hole and in section \ref{secconclusion} we present our final comments and conclusion.

\section{Random walk representation of the heat kernel and quasiperiodicity}
\label{seckernel}
The idea of quantum geometry can be very intuitive. The concept of curvature of a manifold is studied in semiriemannian geometry. In this context the manifold -the spacetime- is smooth, however if nature is quantum at the most fundamental level then the smooth spacetime should be quantized-the notion of quantum geometry. 

Let us describe the idea of spacetime a little more deeply. In general relativity (GR) the spacetime is made of events, these events can be in principle anything: the explosion of a bomb, the hand shake of two friends, the click one do on the mouse, etc. However if we think careful on this definition of spacetime one realizes that something strange happens if we want to describe events with quantum systems as for example with a transition of one level of energy to another in the hydrogen atom\footnote{Similar ideas are consider in \cite{vi14}}. Clearly this happens because the idea of spacetime described in standard GR books is classical. But then we face a conceptual problem similar to the one of the measurement problem in quantum mechanics, as the question where lies the boundary between the machine which measures and the system under study. 

Just to put things on perspective we are aiming to construct something like
\begin{equation}
G(\gamma)=\int_{\gamma}D\sigma e^{-S(\sigma)},
\end{equation}   
where $\sigma$ is an hypersurface, $\gamma$ is its boundary and $S$ is the action of the system. Most of the actions constructed so far are geometric, but since we are constructing a theory more general than the ones we have at the moment we will not attach from the beginning to geometric actions. Clearly the two challenges in this aim are the construction of the measure $D\sigma$ and the action $S(\sigma)$. We are working with a kind of Euclidean action, which is not a limitation because we want to have spacetime emergent in our model . Also it is worthwhile to mention that from $G(\gamma)$ we expect to obtain a kind of generalized partition function.

\subsection{Some mathematical tools from quasicrystals}

A quasicrystal is an object that has order but not periodicity. The mathematics to study these object is very rich and very well developed, see for instance \cite{baa13}, \cite{se96} and \cite{ja89} just to mention a few references.

In the study of quasicrystals, quasiperiodic functions are relevant. The idea of this work is to construct a spacetime foam-like model. First in one dimension with quasiperiodic functions, as the ones shown in \cite{baa13}. Later we will introduce stochasticity too. So our quantum geometry will be the result of nonperiodicity and stochasticity.

\subsection{First steps in the construction of Quantum Geometry}

The main idea of this section is to replace the stochastic process such a random walk used for example in \cite{cha01} by a quasiperiodic random process described by a quasiparticle. However as a first step in this direction we will study the quasiperiodic process described by a quasiperiodic function as the one given in \cite{baa13}.

\subsubsection{The random walk representation of the heat kernel and quasiperiodicity}

In order to have an idea on how to implement quasiperiodicity in the models of quantum geometry let us study some aspects of the random walk representation of the heat kernel associated with the diffusion equation when we introduce quasiperiodicity in the model. This discussion rather than new is pedagogical, for more details see \cite{am97}.

Let $\Delta$ denote the Laplace operator in $R^{d}$. The solution to the difussion (or heat) equation in $R^{d}$
\begin{equation}
\frac{\partial\varphi}{\partial t}=\frac{1}{2}\Delta\varphi,
\end{equation}
with the initial condition $\varphi(x,0)=\varphi_{0}(x)$ is given by
\begin{equation}
\varphi(y,t)=\frac{1}{(2\pi t)^{d/2}}\int_{R^{d}}dxe^{-\frac{\left|x-y\right|^{2}}{2t}}\varphi_{0}(x).
\end{equation}
The function $\varphi_{0}(x)$ is interpreted as the initial distribution of particles at time $t=0$, and $\left|x-y\right|$ denote the Euclidean distance between $x$ and $y$ in $R^{d}$.

The kernel $K_{t}(x,y)$ of the operator $e^{\frac{t}{2}\Delta}$, is called the \textit{heat kernel} and is given by
\begin{equation}
K_{t}(x,y)=\frac{1}{(2\pi t)^{d/2}}e^{-\frac{\left|x-y\right|^{2}}{2t}},
\end{equation}
and represents the probability density of finding the particle at $y$ at time $t$ given its location at $x$ at time $0$. From the simigroup property 
\begin{equation}
e^{(t+s)\Delta}=e^{t\Delta}e^{s\Delta},
\end{equation}
for $s$, $t$$\geq 0$ we have
\begin{equation}
K_{t}(x,y)=\int dx_{1}...dx_{N-1}K_{t/N}(x_{N},x_{N-1})...K_{t/N}(x_{1},x_{0})
\end{equation}
for each $N\geq 1$, where we have set $x_{0}=x$ and $x_{N}=y$.

There is an obvious one-to-one correspondence between configurations $(x_{1},...,x_{N-1})$ and parametrized piecewise linear paths $\omega: [0,t]\rightarrow R^{d}$ from $x$ to $y$ consisting of line segments $[x_{0},x_{1}]$, $[x_{1},x_{2}]$,...,$[x_{N-1},x_{N}]$, such that the segment $[x_{i-1},x_{i}]$ is parametrized linearly by $s$$\in$$[\frac{i-1}{N}t,\frac{i}{N}t]$. We denote the collection of all such paths by $\Omega_{N,t}(x,y)$. Hence we may consider
\begin{equation}
D_{t}^{N}\omega=(2\pi\frac{t}{N})^{-\frac{d}{2}N}dx_{1}...dx_{N-1}
\end{equation}
as a measure on the finite dimensional space $\Omega_{N,t}(x,y)$.

Noting that
\begin{equation}
\sum_{i=1}^{N}\frac{\left|x_{i}-x_{i-1}\right|^{2}}{t/N}=\sum_{i=1}^{N}\frac{t}{N}(\frac{\left|x_{i}-x_{i-1}\right|}{t/N})^{2}=\int_{0}^{t}\left|\dot{\omega}(s)\right|^{2}ds,
\end{equation}
where $\dot{\omega}$ is the piecewise constant velocity of the trajectory $\omega$, hence we can write
\begin{equation}
K_{t}(x,y)=\int_{(x,y)}D_{t}^{N}\omega\exp(-\frac{1}{2}\int_{0}^{t}\left|\dot{\omega}(s)\right|^{2}ds),
\end{equation}
where the suffix $(x,y)$ indicates that paths are restricted to go from $x$ to $y$. We refer to this equation as a \textit{random walk representation of} $K_{t}(x,y)$ on $\Omega_{N,t}(x,y)$.

More generally, given an action functional $S$ on a piecewise linear parametrized paths, we call the equation 
\begin{equation}
H_{t}^{N}(x,y)=\int_{(x,y)}D_{t}^{N}\omega e^{-S(\omega)}
\end{equation}
a random walk representation of the kernel $H_{t}^{N}(x,y)$ on $\Omega_{N,t}(x,y)$.

In is clear from the expressions for the heat kernel that the introduction of quasiperiodicity in the partition of the intervals will bring new features that is worth to be investigated.

\subsubsection{Quasiperiodic Brownian movement}

As a warm up let us write down the transition probability when a particle follows a quasiperiodic Brownian motion. The quasiperiodicity can be introduced with a concrete function such as \cite{baa13}
\begin{equation}
x(\tau)=\cos(2\pi\tau)+\cos(2\pi\alpha\tau),
\end{equation}  
where $\alpha$ is a irrational number. Now if we interpret this function as given the position of a particle after a time $\tau$ then according to the well know evolution of this movement we have that the probability of being at $(\tau, x(\tau))$ if at $\tau=0$ it was at $x=0$ is given by \cite{cha01}
\begin{equation}
W(x(\tau),\tau;0,0)=\frac{1}{\sqrt{4\pi D\tau}}\exp\{-\frac{x^2(\tau)}{4D\tau}\}.
\end{equation}

\section{Random walk on a Fibonacci chain}
\label{secwalk}
In this section we will review the general random walk procedure and then apply it to the Fibonacci chain as preparation for studying random walks on more involved geometries. We will restrict ourselves to the random walk in one dimension.

Let us suppose we have a random walker which can move on a line. Let us denote its position as $X_{n}$ which can be any integer. Now suppose this walker can move to the left or to the right with equal probability\footnote{The probabilities can be different, for example in the case we have a slope.} $1/2$ and the length of the step being $l$. We would like to know the probability that the walker is $n_{R}$ steps to the right and $n_{L}$ steps to the left. And also the probability of being a distance $m$ from the origin after $n_{R}$ steps to the right. This problem is discussed in \cite{re65} and now we will give the solution.

Since each step has length $l$ the location of the walker must be of the form $x=ml$ where $m$ is an integer. A question of interest is the following: after N steps what is the probability of being located at the position $x=ml$? 

One can readily generalize this one-dimensional problem to more dimensions. One again asks for the probability that after $N$ steps the walker is located at certain distance from the origin, however this distance is no longer of the form $ml$. Also on higher dimensions we add vectors of equal length in random directions and then we ask the probability of the resultant vector being in certain direction and certain magnitude. This is exemplified by the following two examples: 

a) Magnetism: An atom has spin $1/2$ and magnetic moment $\mu$; in accordance with quantum mechanics, its spin can point up or down with respect to certain direction. If both possibilities are equally likely, what is the net total magnetic moment of $N$ such atoms?

b) Diffusion of a molecule in a gas: A given molecule travels in three dimensions a mean distance $l$ between collisions with other molecules. How far is likely to have gone after N collisions?

The random walk problem illustrates some very fundamental results of probability theory.The techniques used in the study of this problem are powerful and basic, and recur again and again in statistical physics.

After a total of $N$ steps of length $l$ the particle is located at $x=ml$ where $-N\leq m\leq N$. We want to calculate the probability $P_{N}(m)$
of finding the particle at $x=ml$ after $N$ steps. The total number of steps is $N=n_{L}+n_{R}$ and the net displacement in units of $l$ is given by $m=n_{R}-n_{L}$. If it is known that in some sequence of $N$ steps the particle has taken $n_{R}$ steps to the right, then its net displacement from the origin is determined. Indeed
\begin{equation}
m=n_{R}-n_{L}=n_{R}-(N-n_{R})=2n_{R}-N.
\end{equation} 
This shows that if $N$ is odd then $m$ is odd and if $N$ is even then $m$ is even too.

A fundamental assumption is that successive steps are statistically independent. Thus we can assert simply that, irrespective of past history, each step is characterized by the respective probabilities
\begin{equation}
p=\texttt{probability that the step is to the right}
\end{equation}
\begin{equation}
q=1-p=\texttt{probability that the step is to the left}.
\end{equation}
Now, the probability of a given sequence of $n_{R}$ steps to the right and $n_{L}$ step to the left is given simply by multiplying the probability of each step and is given by
\begin{equation}
\underbrace{pp...p}_{n_{R}\texttt{factors}}\underbrace{qq...q}_{n_{L}\texttt{factors}}=p^{n_{R}}q^{n_{L}}.
\end{equation}
There are several ways to take $n_{R}$ steps to the right and $n_{L}$ steps to the left in $N$ steps. By known combinatorial calculus this number is given by
\begin{equation}
\frac{N!}{n_{R}!n_{L}!}.
\end{equation}
Hence the probability $W_{N}(n_{R})$ of taking $n_{R}$ steps to the right and $n_{L}=N-n_{R}$ steps to the left in $N$ total steps is given by
\begin{equation}
W_{N}(n_{R})=\frac{N!}{n_{R}!n_{L}!}p^{n_{R}}q^{n_{L}}.
\end{equation}
This probability function is known as the binomial distribution. The reason is because the binomial expansion is given by
\begin{equation}
(p+q)^{N}=\sum_{n=0}^{N}\frac{N!}{n!(N-n!)}p^{n}q^{N-n}.
\end{equation}
We already pointed out that if we know that the particle has made $n_{R}$ steps to the right in $N$ total steps then we know its net displacement $m$. Then the probability of the particle being at $m$ after $N$ steps is
\begin{equation}
P_{N}(m)=W_{N}(n_{R}).
\end{equation}
We find explicitly that
\begin{equation}
n_{R}=\frac{1}{2}(N+m)\hspace{1cm}n_{L}=\frac{1}{2}(N-m).
\end{equation}
Hence, in general we have that
\begin{equation}
P_{N}(m)=\frac{N!}{((N+m)/2)!((N-m)/2)!}p^{(N+m)/2}(1-p)^{(N-m)/2}.
\end{equation}
In the special case when $p=q=1/2$ then
\begin{equation}
P_{N}(m)=\frac{N!}{((N+m)/2)!((N-m)/2)!}(1/2)^{N}.
\end{equation}

\subsection{Generalized random walk and the Fibonacci chain case}

Now we will study the generalized random walk. The random walk can be studied in several dimensions, and we will do this up to a certain point and later we will focus on one dimension and finally on the random walk on the Fibonacci chain. In this subsection we mainly follow \cite{hu95}. 

Let $P_{n}(\textbf{r})$ denote the probability density function for the position $\textbf{R}_{n}$ of a random walker, after $n$ steps have been made. In other words, the probability that the vector $\textbf{R}_{n}$ lies in an infinitesimal neighbourhood of volume $\delta V$ centered on $\textbf{r}$ is $P_{n}(\textbf{r})\delta V$. The steps are to be taken independent random variables and we write $p_{n}(\textbf{r})$ for the probability density function for the displacement of the $n$th step. Then the evolution of the walk is governed by the equation
\begin{equation}
P_{n+1}(\textbf{r})=\int p_{n+1}(\textbf{r}-\textbf{r}')P_{n}(\textbf{r}')d^{d}\textbf{r}',
\end{equation} 
where the integral is over all of $d$-dimensional space. This equation is an immediate consequence of the independence of the steps.

It is important to note that, by hypothesis, the probability density function for a transition from $\textbf{r}'$ to $\textbf{r}$ is a function of $\textbf{r}-\textbf{r}'$ only, and not on $\textbf{r}$ and $\textbf{r}'$ separately. In other words, the process is translationally invariant; it is the relative position, not absolute location, which matters. The analysis become much harder when $p_{n+1}(\textbf{r}-\textbf{r}')$ must be replaced by $p_{n+1}(\textbf{r},\textbf{r}')$.

The assumed translational invariance ensures that the formal solution of the problem is easily constructed using Fourier transform. The Fourier transform $\widetilde{p}(q)$ of a function $p(x)$ is defined as
\begin{equation}
\widetilde{p}(q)=\int_{-\infty}^{\infty}e^{iqx}p(x)dx.
\end{equation}
Under appropriate restrictions on the function $p(x)$, there exist an inversion formula:
\begin{equation}
p(x)=\frac{1}{2\pi}\int_{-\infty}^{\infty}e^{-iqx}\widetilde{p}(q)dq.
\end{equation}    

These equations are easily generalized to $d$ dimensions. The Fourier transform becomes
\begin{equation}
\widetilde{p}(\textbf{q})=\int_{-\infty}^{\infty}e^{i\textbf{q}\cdot\textbf{r}}p(\textbf{r})d^{d}\textbf{r},
\end{equation}
where $d^{d}\textbf{r}$ denotes de $d$-dimensional volume element and the integral is taken over all of $d$-dimensional space. Similarly the inversion formula becomes
\begin{equation}
p(\textbf{r})=\frac{1}{(2\pi)^{d}}\int_{-\infty}^{\infty}e^{-i\textbf{q}\cdot\textbf{r}}\widetilde{p}(\textbf{q})d^{d}\textbf{q}.
\end{equation}
The convolution theorem for the Fourier transform states that under modest restrictions on $g$ and $h$
\begin{equation}
k(x)=\int_{-\infty}^{\infty}g(x-x')h(x')dx'\hspace{0.5cm}\texttt{corresponds to}\hspace{0.5cm}\widetilde{k}(q)=\widetilde{g}(q)\widetilde{h}(q).
\end{equation}
The generalization of the convolution theorem to $d$ dimensions is straightforward:
\begin{equation}
k(\textbf{r})=\int_{-\infty}^{\infty}g(\textbf{r}-\textbf{r}')h(\textbf{r}')d^{d}\textbf{r}'\hspace{0.5cm}\texttt{corresponds to}\hspace{0.5cm}\widetilde{k}(\textbf{q})=\widetilde{g}(\textbf{q})\widetilde{h}(\textbf{q}).
\end{equation}
Taking the Fourier transform of our equation for the probabilities we have that
\begin{equation}
\widetilde{P}_{n+1}(\textbf{q})=\widetilde{p}_{n+1}(\textbf{q})\widetilde{P}_{n}(\textbf{q}).
\end{equation}
With $P_{0}(\textbf{r})$ the probability density function for the initial position of the walker, and $\widetilde{P}_{0}(\textbf{q})$ its Fourier transform, we have that
\begin{equation}
\widetilde{P}_{n}(\textbf{q})=\widetilde{P}_{0}(\textbf{q})\prod_{j=1}^{n}\widetilde{p}_{j}(\textbf{q}).
\end{equation}
Taking the inverse Fourier transform of both sides of this equation, we find the solution for the probability density function for the position after $n$ steps:
\begin{equation}
P_{n}(\textbf{r})=\frac{1}{(2\pi)^{d}}\int e^{-i\textbf{q}\cdot\textbf{r}}\widetilde{P}_{0}(\textbf{q})\prod_{j=1}^{n}\widetilde{p}_{j}(\textbf{q})d^{d}\textbf{q}.
\end{equation}
When all steps have the same probability density function $p(\textbf{r})$ and the walk is taken to commence at the origin of coordinates, so that
\begin{equation}
P_{0}(\textbf{r})=\delta(\textbf{r})\hspace{1cm}\widetilde{P}_{0}(\textbf{q})=1,
\end{equation}
then we have
\begin{equation}
P_{n}(\textbf{r})=\frac{1}{(2\pi)^{d}}\int e^{-i\textbf{q}\cdot\textbf{r}}\widetilde{p}(\textbf{q})^{n}d^{d}\textbf{q}.
\end{equation}
There are very few cases in which this integral can be evaluated in terms of elementary functions. However, much useful information can still be extracted.

Now we will see one of the cases where this integral can be reduced a elementary functions. For a random walk in one dimension with different length steps we have that
\begin{equation}
p(x;l_{n})=\frac{1}{2}(\delta(x-l_{n})+\delta(x+l_{n})).
\end{equation}
Using that
\begin{equation}
\delta(x-l_{n})=\frac{1}{2\pi}\int_{-\infty}^{\infty}dke^{ik(x-l_{n})},
\end{equation}
then we have that
\begin{equation}
\widetilde{p}(q)=\int_{-\infty}^{\infty}e^{iqx}p(x)dx=\frac{1}{2}(e^{iql_{n}}+e^{-iql_{n}})=\cos(ql_{n}).
\end{equation}
Hence
\begin{equation}
P_{n}(x;l_{n})=\frac{1}{2\pi}\int e^{-iqx}\cos^{n}(ql_{n})dq.
\end{equation}
In the case of the Fibonacci sequence we have
\begin{equation}
l_{n+1}=l_{n}+l_{n-1}\hspace{0.5cm}with\hspace{0.5cm}l_{0}=0, l_{1}=1.
\end{equation}
So in this case we can solve the problem completely.

There is a subtlety with this expression for the probability, it diverges. The problem is that we are dealing with distributions and classical analysis does not work here. So we have to use the distribution theory. From p. 63 of \cite{hu95} we know that the correct expression for the probability is
\begin{equation}
Pr\left\{X_{n}=ll_{n}\right\}=\frac{l_{n}}{2\pi}\int_{-\pi/l_{n}}^{\pi/l_{n}}e^{-ill_{n}\xi}\cos^{n}(l_{n}\xi)d\xi,
\end{equation} 
where $l\in Z$. It is interesting that if we change variables as $l_{n}\xi=k$ then
\begin{equation}
Pr\left\{X_{n}=ll_{n}\right\}=\frac{1}{2\pi}\int_{-\pi}^{\pi}e^{-ilk}\cos^{n}kdk,
\end{equation}
and there is no dependence of $l_{n}$ in the integral.

\subsection{The random walk in a two dimensional Fibonacci lattice}

Now let us consider a infinite two dimensional Fibonacci lattice. Then in this case the probability density is given by
\begin{equation}
p(x,y;l_{nx},l_{ny})=\frac{1}{4}(\delta(x-l_{nx})+\delta(x+l_{nx})+\delta(y-l_{ny})+\delta(y+l_{ny})).
\end{equation} 
Then following the one dimensional case we have that in the present case the probability function is given by
\begin{equation}
P_{n}(x,y;l_{nx},l_{ny})=\frac{1}{8\pi}(\int e^{-iqx}\cos^{n}(ql_{nx})dq+\int e^{-ipx}\cos^{n}(pl_{ny})dp).
\end{equation}
Here $q$ and $p$ are variables in the Fourier space and $l_{nx}$ and $l_{ny}$ are Fibonacci numbers.

Making the corresponding manipulations we did in the 1-dimensional Fibonacci sequence, now we obtain in this case
\begin{equation}
Pr\left\{X_{n}=ll_{nx},Y_{n}=ml_{ny}\right\}=\frac{1}{8\pi}(\int_{-\pi}^{\pi}e^{-ilk}\cos^{n}kdk+\int_{-\pi}^{\pi}e^{-imk}\cos^{n}kdk),
\end{equation}
where $l,m\in Z$.

\section{The entropy of the Fibonacci chain}
\label{secentropy}
One of the main object in our approach is the a kind of partition function which in certain limit should be reducible to the Einstein-Hilbert action and in other limit to the partition function of quantum statistical mechanics. In order to construct this partition function we will follow the ideas explained in \cite{am97}, \cite{va98} and \cite{fe10}.

Let us give a simple example of the kind of things we are working with. One possible action for a piecewise constant path is \cite{am97}
\begin{equation}
A=\tilde{\beta}\sum_{i=1}^{n}\left|x_{i}-x_{i-1}\right|,
\end{equation}
where we will suppose that $\tilde{\beta}$ is a generalized inverse of the temperature. Then the partition function\footnote{Here we are thinking the action as an effective action which coincides at zero loops with the classical action.} associated with this action is
\begin{equation}
Z=e^{-\tilde{\beta}\sum_{i=1}^{n}\left|x_{i}-x_{i-1}\right|}.
\end{equation} 
The energy associated with this partition function is
\begin{equation}
E=-\frac{\partial}{\partial\tilde{\beta}}\ln Z=\sum_{i=1}^{n}\left|x_{i}-x_{i-1}\right|
\end{equation}
and the entropy is
\begin{equation}
S=E+\ln Z=(1-\tilde{\beta})\sum_{i=1}^{n}\left|x_{i}-x_{i-1}\right|.
\end{equation}
Here the $x_{i}$'s are an homogeneous partition of the path. In this sense it is a periodic partition. It is clear that if now we assume that the $x_{i}$'s are quasiperiodic then the entropy will change. It is not difficult to imagine how hard it would be to solve if instead of having a one-dimensional path we have a surface or a volume. It could be interesting to compare the entropy $S$ with the entropy of a elastic string. If we want the discrete action to go to the continuous action as the size of the partition goes to zero then $\tilde{\beta}$ should depend on the size of the partition function \cite{am97}. Then clearly in this case if we choose a quasicrystalline partition then the entropy and other thermodynamical quantities will be impacted.

\subsection{Partition function and entropy of the Fibonacci chain}

If we consider the Fibonacci chain in 1-dimension, we can define a partition function as
\begin{equation}
\textsl{Z}=\sum_{n}Pr\left\{X_{n}=ll_{n}\right\}.
\end{equation}

Analogously in the 2-dimensional case we have then
\begin{equation}
\textsl{Z}=\sum_{n}Pr\left\{X_{n}=ll_{nx},Y_{n}=ml_{ny}\right\}.
\end{equation}
If these definitions are correct, then it is a matter of brute force to calculate the analogous of thermodynamical quantities.

For example let us do this for the 1-dimensional Fibonacci chain. In this case we would have that the entropy is given by
\begin{equation}
S=F(l_{n})\left\langle X_{n}\right\rangle+\ln Z,
\end{equation}
where $F(l_{n})$ is a function which we should determine using plausible arguments and $\left\langle X_{n}\right\rangle$ is the expectation value of $X_{n}$. 

Analogously, in the two dimensional case we have
\begin{equation}
S=G(l_{nx},l_{ny})\left\langle X_{n}, Y_{n}\right\rangle+\ln Z,
\end{equation}
where $G(l_{nx},l_{ny})$ is a function we have to propose. For example, if we agree that with a new step there is an increasing of information then these functions should be decreasing functions of the lengths. This entropy can be considered an entropy where randomness and geometry are mingled and can be useful in a theory of quantum geometry.

\section{On the Euclidean action and the Boltzman factor}
\label{secbtz}
One of our goals is to construct an object that in one limit gives the General Relativity action (classical and quantum) and in the other side gives the quantum mechanical statistics partition function.

In the book \cite{hu10} Huang says that it is a deep mystery of physics that the Hamiltonian operator appears in the evolution operator in quantum mechanics and in the partition function in quantum statistical mechanics:
\begin{equation}
e^{-it\hat{H}}\hspace{3cm}e^{-\beta\hat{H}}.
\end{equation}
Here $\beta=\frac{\kappa}{T}$, with $T$ the temperature and $\kappa$ the Boltzman's constant. If we make $t=-i\tau$, where $\tau$ is real and periodic with period of $\beta$ then both expressions become the same. 

The purpose of this section is to comment on this deep mystery and to try to elucidate, at least partially, why this occurs.

We think this discussion is important since important results such as the entropy of black holes in euclidean quantum gravity \cite{haw93} uses this deep mystery.

We discuss next how the Boltzman factor is related to the action, and 
how the entropy of the BTZ black hole is obtained in Euclidean Quantum Gravity.

\subsection{On the action and the Boltzman factor}

It is interesting to note that the action $S$ of a system appears in the path integral \cite{cha01}, \cite{fe10}, the partition function \cite{am97} and the Hamilton-Jacobi equation \cite{fe03}. Also it is interesting the similarity between the Boltzman factor and the normal distribution. Let us elaborate on these two ideas. 

In the Euclidean setting we have the path integral
\begin{equation}
A=\int Dx e^{\frac{-S}{\hbar}}.
\end{equation}
Whereas the Boltzman factor is
\begin{equation}
B_{i}=e^{\frac{-E_{i}}{kT}}.
\end{equation}
We know the action has units of energy times time. So if we multiply in the Boltzman factor the energy and the $kT$ term by some time we have an term with units of action. Now the partition function is 
\begin{equation}
Z=\sum_{i} B_{i}=\sum_{i}e^{\frac{-E_{i}}{kT}}.
\end{equation}
The similarity with the path integral is obvious. Now the normal distribution has the following form
\begin{equation}
N=ne^{-\frac{x^2}{D}}.
\end{equation}
Clearly if we multiply the square term by one over time square, and also the D, then we have energy units. In one step further we can have a kind of action in the normal distribution. Now, the Boltzman distribution is ubiquitous in statistical mechanics and so is the normal distribution in several natural processes. From this point of view the normal distribution is analogous to the expression of the effective action\footnote{See for example \cite{to09} where the relationship between the path integral and the effective action is displayed.}. So one may wonder if there is a deep connection between these three expressions. One might wonder if we can make up a mechanical toy model where in one side one has the normal distribution and on the other end tha path integral and in the middle the partition function obtained from the Boltzman factor. If we impose a periodicity in the Euclidean amplitude A then with the correct units we have the well know temperature of Black Holes. This periodicity when seen from a discrete system can be related to the Poincar\'{e} recurrence theorem.

This toy model seems to be relevant for the unification physics since in one hand one has a discrete system (similar to a quantum geometry) and in the other hand a continuous system (module some metric issues) similar to a topological quantum field theory.

It is also interesting that the action appears in the Hamilton-Jacobi equation whose quantum limit is the Schrodinger equation and it can branch to classical mechanics, gravitational physics and electromagnetic theory.

Just to finish this section we note that the Lagrangian is given as
\begin{equation}
L=E-V,
\end{equation}
where $E$ is the energy and $V$ the potential (energy). Hence we see that the Lagrangian is a kind of generalized energy. The action is 
\begin{equation}
S=\int Ldt.
\end{equation}
Hence when we make $t$ imaginary and periodic, with the correct period in, for example, black holes then everything about time drops and we have the partition function of statistical mechanics. It is as if there were hidden a symmetry related with time. Here we have taken the simplest Lagrangian however it is not difficult to see that for example the scalar field the situation is very similar.

The above discussion makes clear why the Euclidean path integral coincides with the partition function when the time is periodic, in some sense the partition function is hidden in the path integral.

\subsection{On the entropy of black holes}

As one example of some of the ideas presented in the previous section now we will explain how the entropy of some black hole can be obtained using the effective action.

It is well know, see for example \cite{kl16}, that at zero order the effective action $\Gamma[\Phi]$ coincides with the classical action $A[\Phi]$ evaluated on the mean field $\Phi$. 

The evaluation of the black hole entropy of the Kerr black hole can be consulted \cite{haw93}, and now we will show how the entropy for the BTZ black hole is obtained.

We follow mainly \cite{ku05}. The Euclidean action of the BTZ black hole is
\begin{equation}
I_{E}=\beta M-\frac{A}{4G}.
\end{equation}
Then the partition function is 
\begin{equation}
Z_{BTZ}(T)=\exp^{\frac{(\pi l)^2T}{2G}},
\end{equation}
where $l$ is the AdS radius. The expectation value of the energy is
\begin{equation}
E_{BTZ}=-\frac{\partial}{\partial\beta}\ln Z=M.
\end{equation}
Whereas the entropy is given by
\begin{equation}
S_{BTZ}=\beta E_{BTZ}+\ln Z_{BTZ}=4\pi r_{+}=\frac{A}{4G}.
\end{equation}
Which is the result one expects on the grounds of Beckenstein ideas on entropy of black holes.

It is interesting to note that there are at least three other values of the BTZ black hole entropy obtained in \cite{is08}, \cite{ich95} and \cite{ka14}. In the first in loop quantum qravity, the second in standard statistical field theory and in the third in the brick wall model. In the first two models it does not coincide with the value given in \cite{ku05} whereas in the brick wall model it coincides with \cite{ku05}.

\section{Final comments and conclusions}
\label{secconclusion}
In this work we have explored how randomness, geometry and aperiodicity mingle in a coherent way and have the potential of giving interesting results in computation of physical observables. In particular this work is a pioneer in the study of aperiodicity in aspects of gravity. 

It is clear that the introduction of aperiodicity in the framework of quantum gravity would give substantially different results compare with the standard approaches. Hence it would be interesting in the future to do something similar with other quantum gravity approaches. 

From the considerations in this work it is clear that our approach is closer to the standard path integral approach than to the Hilbert space framework. In this sense it would be interesting if with our approach we can recover the well known results from Euclidean quantum gravity as explained in \cite{haw93}, \cite{ca98}.

It is interesting to note the following: The result of \cite{ku05} is classical, although using a quantum framework, the result of \cite{is08} is quantum but it does not give the expected result, the result of \cite{ich95} is semiclassical and gives a close result to the one expected, and finally the result of \cite{ka14} is quantum and gives the expected result but the entropy is of the scalar field living on the BTZ black hole. Hence there is no a consensus about this entropy. Just to finish up we note that the temperature of a black hole does not make sense without a field living on it, so, after all the brick wall model could be the one closer to the origin of the BTZ black hole entropy.\vspace{1cm}

Acknowledgments: This work is fully sponsored by Quantum Gravity Research.

\end{document}